\documentstyle[12pt,graphicx,caption2]{article}

\begin{document}

\title{Limit on the neutron-antineutron transitions in the alternative models}
\author{V.I.Nazaruk\\
Institute for Nuclear Research of RAS, 60th October\\
Anniversary Prospect 7a, 117312 Moscow, Russia.}

\date{}
\maketitle
\bigskip

\begin{abstract}
We summarize our study of the neutron-antineutron transition in medium and nuclei in the framework of the field-theoretical approach. The models based on the diagram technique for direct reactions, potential description of antineutron-medium interaction are considered as well. The lower limit on the free-space neutron-antineutron oscillation time $\tau $ is in the range: $10^{16}\; {\rm yr}>\tau >1.2\cdot 10^{9}\; {\rm s}$.
\end{abstract}

\vspace{5mm}

\vspace{1cm}

\vspace{1cm}

*E-mail: nazaruk@inr.ru

\newpage
\setcounter{equation}{0}
\section{Introduction} 
We consider the limits on the free-space $n\bar{n}$ oscillation time $\tau $ obtained from the nuclear stability lifetime by means of alternative models.
Let us consider the $n\bar{n}$ transitions in a medium and nuclei followed by annihilation:
\begin{equation}
n\rightarrow \bar{n}\rightarrow M,
\end{equation}
here $M$ are the annihilation mesons.

As far back as 1992, the model of $n\bar{n}$ transitions in light nuclei based on diagram technique for direct reactions (later on referred to as the model 1) has been published [1,2]. In 1996 this calculation was repeated for deuteron [3]. However, in [4] we abandoned this model for reasons given below. The model based on field-theoretical approach with finite time interval (model 2) has been proposed [4-6].

The models 1 and 2 give radically different results. This is because the $n\bar{n}$ transition in nuclei is extremely sensitive to the details of the model and so we focus on the physics of the problem. Recently, the calculation of [1-3] was repeated [7]. So the model 1 we consider by the example of Ref. [7]. The main purpose of this paper is the analysis and comparison of the models 1 and 2 and elucidation of the reasons of the result sensitiveness to the details of the model. 

We compare the model based on the field-theoretical approach (section 2) with the model based on diagram technique for direct reactions (section 3) and show that the latter model is unsuitable for the problem under study. Also it is shown that interpretation of a number of important points of the above-mentioned models given in [7] is wrong or irrelevant (sections 2-4). In section 5 we touch on the model based on potential description of $\bar{n}$-medium interaction (potential model). Section 6 contains the conclusion.

\section{Field-theoretical approach}
Model 2 describes both $n\bar{n}$ transition in the medium [4,5] (Fig. 1a) and nuclei [8] (Fig. 1b). Fig. 1 is the process model in the coordinate representation (graphic presentation). It corresponds to the standard formulation of the problem: the $|in>$-state is the eigenfunctions of unperturbed Hamiltonian $H_0$. In the case of the process shown in Fig. 1a, this is the neutron plane wave. In the case of Fig. 1b, this is the wave function of bound state [8,9]. For the nucleus in the initial state we take the one-particle shell model. 

We consider the process shown in Fig. 1b. The initial neutron state $n(x) $ is given by

\begin{eqnarray}
(i\partial_t-H_0)n(x)=0,\nonumber\\
H_0=-\nabla^2/2m+U(x),
\end{eqnarray}
where $U(x) $ is the self-consistent neutron potential. The eigenfunctions $n_j({\bf x})$ of the unperturbed Hamiltonian $H_0n_j({\bf x})=\epsilon _jn_j({\bf x})$ ($\epsilon _j$ are the energies of stationary states) form the complete orthogonal set.

\begin{figure}[!h]
%  \reflectbox{\includegraphics[height=.25\textheight]{golfer}}
  {\includegraphics[height=.2\textheight]{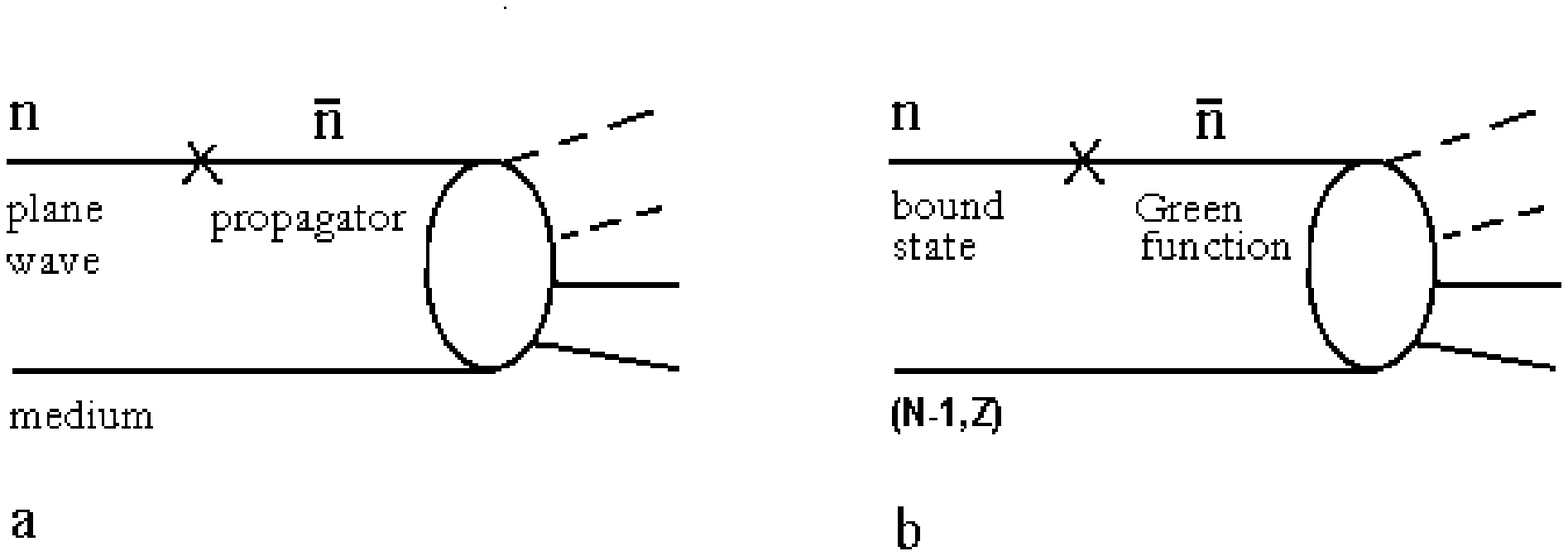}}
  \caption{$n\bar{n}$ transition in the medium ({\bf a}) and nuclei ({\bf b}) followed by annihilation.}
\end{figure}

The Green function $G$ of Eq. (2) is defined as
\begin{equation}
[i\frac{\partial}{\partial t'}-H_0({\bf x}')]G(x',x)=\delta ^3({\bf x}'-{\bf x}) 
\delta (t'-t).
\end{equation}

The density of the interaction Hamiltonian is
\begin{eqnarray}
{\cal H}_I={\cal H}_{n\bar{n}}+{\cal H},\nonumber\\
{\cal H}_{n\bar{n}}=\epsilon \bar{\Psi }_{\bar{n}}\Psi _n+H.c.,
\end{eqnarray}
$H(t)=\int d^3x{\cal H}(x)$. Here ${\cal H}_{n\bar{n}}$ and ${\cal H}$ are the densities of the Hamiltonians of the $n\bar{n}$ transition [10] and the $\bar{n}$-nuclear interaction, respectively; $\epsilon $ is a small parameter with $\epsilon =1/\tau $, where $\tau $ is the free-space $n\bar{n}$ 
oscillation time; $\Psi _n$ and $\Psi _{\bar{n}}$ are the operators of the neutron and antineutron fields; $m_n=m_{\bar{n}}=m$. 

\subsection{$n\bar{n}$ transition in the medium}
Consider now the process shown in Fig. 1a. Put $U(x)=U_n=$const. Then 
\begin{equation}
n(x)=\Omega ^{-1/2}\exp (-ipx), 
\end{equation}
$p=(\epsilon _n,{\bf p}_n)$, $\epsilon _n={\bf p}_n^2/2m+U_n$. Equation (3) has 
the simple solution
\begin{equation}
G=\frac{1}{\epsilon _{\bar{n}}-{\bf p}_{\bar{n}}^2/2m-U_n+i0}\sim \frac{1}{0}.
\end{equation} 
$G\sim 1/0$ since $\epsilon _{\bar{n}}=\epsilon _n$, ${\bf p}_{\bar{n}}={\bf p}_n$. This is an infrared singularity conditioned by zero momentum transfer in the $n\bar{n}$ transition vertex. For the process shown in Fig. 1b the amplitude is singular for the same reason. (In the case of realistic potentials $U(x)$ the solution in analytical form is unavailable. To stress this circumstance for antineutrons in Figs. 1a and 1b the different terms are used.)

In principle, the part of the Hamiltonian ${\cal H} $ can be included in the antineutron Green function [5]. Then the antineutron self-energy $\Sigma $ is generated. In Eq. (6) one should replace
\begin{equation}
G\rightarrow G_d=\frac{1}{\epsilon _{\bar{n}}-{\bf p}_{\bar{n}}^2/2m-U_n-\Sigma +i0}\neq \frac{1}{0}.
\end{equation}

So in the model 2 the antineutron propagator can be bare or dressed. The corresponding results differ radically. If the antineutron propagator is bare  (the antineutron self-energy $\Sigma =0$), the $S$-matrix amplitudes corresponding to Figs. 1a and 1b contain the infrared singularity. To avoid the  singularities, the problem is formulated on the finite time interval $(t,0)$ [6]. Then the matrix element of evolution operator $U(t,0)$ is calculated (see Eq. (13) of Ref. [4]). Eventually for the model 2 with bare propagator the lower limit on the free-space $n\bar{n}$ oscillation time was found to be [4,5] 
\begin{equation}
\tau ^b=10^{16}\; {\rm yr}.
\end{equation}
This value is interpreted as the estimation from above.

The fact that the evolution operator is used is not new [9,11]. Recall that the most part of physical problems is formulated on the finite time interval. The model described above is standard and so it created no questions up to now. However, in [7] we read:

1) "The author of [4-6] tries to reconstruct the space-time picture of the process..."
In reality we do not more than calculate the matrix element of evolution operator.

2) "...new rules seem to be proposed in [4,5] instead of well known Feynman rules."

We calculate the matrix element of evolution operator only. In regard Feynman rules, the problem formulation (process model) in the quantum field theory and nuclear physics differ principally. In particular, there is density-dependence (coordinate-dependence) in nuclear physics. This is absolutely different problems and so the Feynman rules developed in quantum electrodynamics can be only element of the corresponding model of nuclear reaction or decay. This model (more precisely, the model of $n\bar{n}$ transitions in light nuclei based on diagram technique for direct reactions) was published in [1,2] and analyzed in the next section. So the words "instead of well known Feynman rules" mislead the reader at least. 

3) "These new rules should allow reproducing all the well known results of nuclear theory". 

The consideration of the matrix element of evolution operator is interpreted [7] as "new rules". As for the second part of the remark, indeed, all the well known results of nuclear theory have been obtained by means of $S$-operator ($U$-operator).

The "remarks" 1) - 3) are due to the misunderstanding of the authors of [7]. Also we recall that the above-given problem formulation created no questions up to now. If the antineutron self-energy equal to zero, the process amplitude is singular. In this case the {\em calculation} of matrix element is really non-standard [4,5]. However, the pp. 1)-3) are irrelevant to the calculation of matrix element.

For the model 2 with dressed propagator (see Fig. 1a, where the antineutron propapator is dressed) the calculation is standard [5,12]. In this case (model 2 with dressed propagator) the lower limit on the free-space $n\bar{n}$ scillation time is [5,12]
\begin{equation}
\tau ^d=1.2\cdot 10^{9}\; {\rm s}.
\end{equation}

\subsection{$n\bar{n}$ transition in the nuclei}
For the $n\bar{n}$ transition in the nuclei (Fig. 1b) the results are the same [8]: Egs. (8) and (9) for the model 2 with bare and dressed propagator, respectively.

The models with bare and dressed propagators were analyzed and compared in [5,12]. The main purpose of this paper is the analysis of model 1.

\section{Model based on the diagram technique for direct reactions}
Model 1 is shown in Fig. 2. Denote: $A=(N,Z)$, $B=(N-1,Z)$ are the initial and
\begin{figure}[!h]
%  \reflectbox{\includegraphics[height=.2\textheight]{golfer}}
  {\includegraphics[height=.15\textheight]{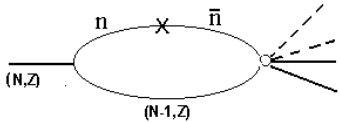}}
  \caption{Model 1 for the $n\bar{n}$ transition in the nuclei followed by annihilation.}
\end{figure}
intermediate nuclei, $M$ is the amplitude of virtual decay $A\rightarrow n+(A-1)$, $M_a^{(n)}$ is the amplitude of $\bar{n}B$ annihilation in $(n)$ mesons, $E_n$ is the pole neutron binding energy; $m$, $m_A$, $m_B$ are the masses of the nucleon and nuclei $A$ and $B$, respectively. The process amplitude is [2]
\begin{equation}
M^{(n)}=-\frac{im^2m_B}{2\pi ^4}\epsilon 
\int d{\bf q}dE \frac{M(q)M_a^{(n)}(m_A)}{({\bf q}^2-2mE-i0)^2[{\bf q}^2+2m_B
(E+E_n)-i0)]}.
\end{equation}
For deuteron the process probability $W_1(t)$ is given by Eqs. (6), (12) and (18) of Ref. [2]:
\begin{eqnarray}
W_1(t)=\Gamma _{d\rightarrow mesons}t,\nonumber\\
\Gamma _{d\rightarrow mesons}=\frac{\epsilon ^2}{6E_n^2}\Gamma _{\bar{n}p},
\end{eqnarray}
where $\Gamma _{\bar{n}p}$ is the $\bar{n}p$ annihilation width. This is final result. At the same time the authors [7] write "There is no final formula for $\Gamma _{d\rightarrow mesons}$ in [2]."

We list the main drawbacks of the above-given model which are essential for the problem under study:

a) The model does not reproduce the $n\bar{n}$ transitions in the medium and vacuum. If $E_n\rightarrow 0$, $W_1$ diverges (see also Eqs. (15) and (17) of Ref. [3]). 

b) Contrary to the model 2, the amplitude (9) cannot be obtained from the Hamiltonian because in the interaction Hamiltonian {\em there is no term} which induces the virtual decay $(N,Z)\rightarrow n+(N-1,Z)$.

c) The model does not contain the infrared singularity for any process including the $n\bar{n}$ transition, whereas it exists for the processes in the medium and vacuum (see [4-6]). This brings up the question: Why? The answer is that for the propagator the infrared singularity cannot be in principle since the particle is virtual: $p_0^2\neq m^2+{\bf p}^2$. Due to this the model is infrared-free. On the other, hand the neutron is in the bound state and should be described by wave function and not the propagator.

d) Since the model is formulated in the momentum representation, it does not describe the coordinate-dependence, in particular the loss of particle intensity due to absorption. Also there is a no the dependence on nuclear density! The model is crude and has very restricted range of applicability. 

e) The non-unitarity of $S$-matrix means, in particular, the probability non-conservation. The model based on the potential description of $\bar{n} $-medium interaction (potential model) is non-unitary. As a result, in this model the effect of final-state absorption (annihilation) acts in the opposite (wrong) direction [5,13]. If we cannot write the Hamiltonian of the model 1, what one can say about unitarity?

We consider the points b) and c). The $n\bar{n}$ transition takes place in the propagator. As the result the model is infrared-free. For the processes with zero momentum transfer this fact is crucial since it changes the functional structure of the amplitude.

On the other hand, the neutron propagator arises owing to the vertex of virtual decay $A\rightarrow n+(A-1)$. However, as pointed out above, in the interaction Hamiltonian {\em there is no term} which induces the virtual decay $A\rightarrow n+(A-1)$. This vertex is the artificial element of the model. It was introduced in order for the neutron (pole particle) state to be separated.

We assert that for the problem under study this scheme is incorrect. The neutron state is described wrongly. The diagram technique for direct reactions has been developed and adapted to the direct type reactions. The term "diagram technique for direct reactions" emphasizes this circumstance. The processes with non-zero momentum transfer are considerably less sensitive to the description of pole particle state. The approach is very handy, useful and simple since it is formulated in the momentum representation. It was applied by us for the calculation of knock-out reactions and $\bar{p}$-nuclear annihilation [14]. The price of simplicity is that its applicability range is {\em restricted}. At the same time, as is seen from p. c), the process under study is extremely sensitive to the description of neutron state. The same is true for the value of antineutron self energy [12]. Besides, the problem is unstable [12].

Finally, since the operator ${\cal H}_{n\bar{n}}$ acts on the neutron, in the model 1 the vertex of virtual decay $A\rightarrow n+(A-1)$ is introduced because one should separate out the neutron state. This scheme is artificial because in the interaction Hamiltonian {\em there is no term} which induces the virtual decay $A\rightarrow n+(A-1)$. The neutron of the nucleus is in the bound state and so it cannot be described by the propagator. More precisely the model containing the vertex of virtual decay is a crude one. It is inapplicable for the problem under study. Alternative method is given by the model 2 which does not contain the above-mentioned vertex. The shell model used in the $|in>$-state of the model 2 has no need of a commentary. 

For above-given reasons we abandoned this model [4]. 

\section{Comparison of the models and discussion}
We continue the comparison of models 1 and 2 by the example of Ref. [7]. In the abstract [7] we read: 

4) "Infrared divergences do not appear within the correct treatment of analytical properties of the amplitudes."

The analytical properties of the amplitudes have nothing to do with the infrared divergences. The infrared divergences do not appear since the model 1 used by authors [7] is, by construction, infrared-free (see section 3). In the model 1 the analytical properties of the amplitudes are related to the calculation of the integral only (which is trivial).

5) The words "we give some general arguments..." are wrong since these arguments result from the model 1.

Figure 1 adduced in [7] determines uniquely the process model (model 1). Equation (8) and subsequent consideration relate to the model 1.

6) The consideration of "energy scale" [7] is nonsense. What is the "energy scale" for nuclear $\beta $-decay, for example?

7) "The reason of suppression is the localization of the neutron inside the nucleus."

To understand that the localization of the neutron inside the nucleus is unrelated to the suppression, it is suffice to compare the nuclear $\beta $-decay and decay of free neutron. The sole possible reason discussed in the literature [13,15] is the antineutron potential. In the models 1 and 2 the antineutron potential is not introduced and above-mentioned mechanism of suppression is inoperative. We also recall that the $n\bar{n}$ transition in the nuclei followed by annihilation is the dynamical, two-step process with the characteristic time $\tau _{ch}\sim 10^{-24}$ s [4,5,12].  

8) "If the infrared divergence takes place for the process of $n\bar{n}$ transitions in nucleus, it would take place also for the nucleus form-factor at zero momentum transfer. But it is well known not to be the case, as we illustrated in this section"

So the authors illustrated that the model 1 is infrared-free and the large part of [7] is devoted to this "problem". However, this trivial fact is well known [9,11]: the infrared divergence cannot take place for virtual particle. 

The errors indicated in pp. 1) - 8) and the fact that the main calculation is the special case of the model proposed by us [1,2] in 1992 are incompatible with the paper title "Critical examination..." and the beginning of the acknowledgements "Present investigation, performed partly with {\em pedagogical} purposes". 

\section{Potential model}
We briefly touch on the model based on potential description of $\bar{n}$-medium interaction. In this model the interaction of $n$ and $\bar{n}$ with the matter (nuclei) is described by the potentials $U_{n},U_{\bar{n}}$. The model can be realized by means of diagram technique or equations of motion (see Refs.[5,13] for further reference):
\begin{eqnarray}
(i\partial_t-H_0)n(x)=\epsilon \bar{n}(x),\nonumber\\
(i\partial_t-H_0-V)\bar{n}(x)=\epsilon n(x),\nonumber\\
H_0=-\nabla^2/2m+U_n,\nonumber\\
V=U_{\bar{n}}-U_n={\rm Re}U_{\bar{n}}-i\Gamma /2-U_n,
\end{eqnarray}
$\bar{n}(0,{\bf x})=0$. Here $U_n$ and $U_{\bar{n}}$ are the potentials of
the $n$ and $\bar{n}$, respectively; $\Gamma $ being the annihilation width of the $\bar{n}$.

With Eq. (12) the matrix element of evolution operator $U_{ii}(t)=1+iT_{ii}(t)=<\!n(0)\!\mid
\!n(t)\!>$ is obtained. The total process width is calculated by means of optical theorem. However, the optical theorem follows from the equation 
\begin{equation}
\sum_{f\neq i}\mid T_{fi}\mid ^2\approx 2ImT_{ii}, 
\end{equation}
$S=1+iT$ which is obtained from the unitarity condition $(SS^+)_{fi}=\delta _{fi}$. But in the potential model the $S$-matrix is {\em essentially} non-unitary $(SS^+)_{fi}\neq \delta _{fi}$ or, what is the same

\begin{equation}
(SS^+)_{fi}=\delta _{fi}+\alpha _{fi},
\end{equation}
$\alpha _{fi}\neq 0$. This equation gives
\begin{equation}
\sum_{f\neq i}\mid T_{fi}\mid ^2\approx 2ImT_{ii}+\alpha _{ii}\neq 2ImT_{ii}
\end{equation}
because $2ImT_{ii}$ is extremely small: $2ImT_{ii}<10^{-31}$ [13]. Instead of Eq. (13), we have Eq. (15). Consequently, optical theorem is inapplicable in this case since it follows from Eq. (13).

As a result, the process (1) probability is $W\sim \Gamma $ [4,5,13], whereas the potential model gives $W\sim 1/\Gamma $ [5,10,13]. In the potential model the effect of final state absorption (annihilation) acts in the opposite (wrong) direction, which tends to the additional suppression of the $n\bar{n}$ transition. Since the annihilation is the main effect which defines the speed of process (1), the potential model should be rejected. This is because the unitarity condition is used for the essentially non-unitary $S$-matrix. The interaction Hamiltonian contains the antineutron optical potential $U_{\bar{n}}$ and ${\rm Im}U_{\bar{n}}$ plays a crucial role. The $S$-matrix should be {\em unitary}. 

The potential model describes only the channel with $\bar{n}$ in the final state [13] when unitarity condition is not used. For the oscillations in the external field the Hamiltonian is hermitian and so there is no similar problem. The above-given remark holds only for the processes (1) and total neutron-antineutron transition probability calculated by means of non-hermitian Hamiltonian. Recall that in the optical model the Schrodinger-type equation (not the system (12)) is considered and unitarization of $S$-matrix takes place.
This problem was considered in Refs.[5,13] in detail.

\section{Conclusion}
In conclusion, for the $n\bar{n}$ transition in medium field-theoretical approach should be used. The process under study is extremely sensitive to the description of neutron state (model 1 or 2). The same is true for the value of antineutron self-energy $\Sigma $. In line with Eqs. (8) and (9), if $\Sigma $ changes in the limits $10\; {\rm MeV}>\Sigma >0\; {\rm MeV}$, the lower limit on the free-space $n\bar{n}$ oscillation time $\tau $ is in the range [8,12]
\begin{equation}
10^{16}\; {\rm yr}>\tau >1.2\cdot 10^{9}\; {\rm s}.
\end{equation}
Usually, only the lower bound for $\tau $ is given ($1.2\cdot 10^9\; {\rm s}$ in our case). We adduce the wide range (16) since $\tau $ depends on $\Sigma $ and the values $\tau >>1.2\cdot 10^{9}\; {\rm s}$, including $\tau =\tau ^b=10^{16}\; {\rm yr}$ are quite realistic.

For the free-space $ab$ oscillations the picture is exactly the same: the $ab$ transition probability is extremely sensitive to the difference of masses $m_a-m_b$. The fact that process amplitude is in the peculiar point (see Fig. 1) is the basic reason why the range (16) is very wide. The values $\tau ^d=1.2\cdot 10^{9}\; {\rm s}$ and $\tau ^b= 10^{16}$ yr are interpreted as the estimations from below (conservative limit) and from above, respectively. The estimation from below $\tau ^d=1.2\cdot 10^{9}\; {\rm s}$ exceeds the restriction given by the Grenoble reactor experiment [16] by a factor of 14 and the lower limit given by potential model [17] by a factor of 5. At the same time the range of uncertainty of $\tau $ is too wide. Further investigations are desirable. The
value of antineutron self-energy $\Sigma $ should be studied first.


\newpage
\begin{thebibliography}{99}
\bibitem{1}
V. I. Nazaruk, Soviet Physics - Lebedev Institute Reports {\bf 11,12}, 29 (1992).
%
\bibitem{2}
V. I. Nazaruk, Yad. Fiz. {\bf 56}, 153 (1993).
%
\bibitem{3}
L. A. Kondratyuk, Pis'ma Zh. Exsp. Theor. Fiz. {\bf 64}, 456 (1996).
%
\bibitem{4}
V. I. Nazaruk, Phys. Rev. C {\bf 58}, R1884 (1998).
%
\bibitem{5}
V. I. Nazaruk, Eur. Phys. J. C {\bf 53}, 573 (2008).
%
\bibitem{6}
V. I. Nazaruk, Phys. Lett. B {\bf 337}, 328 (1994).
%
\bibitem{7}
V. Kopeliovich and I. Potashnikova, Eur. Phys. J. C {\bf 69}, 591 (2010).
%
\bibitem{8}

V. I. Nazaruk, Int. J. of Mod. Phys. E {\bf 20}, 1203 (2011); arXiv: 1004.3192 [hep-ph].
%
\bibitem{9}
J. D. Bjorken and S. D. Drell, {\em Relativistic Quantum Fields} (New York, McGraw-Hill, 1964).
%
\bibitem{10}
K. G. Chetyrkin, M. V. Kazarnovsky, V. A. Kuzmin and M. E. Shaposhnikov, Phys. Lett. B {\bf 99}, 358 (1981).
%
\bibitem{11}
C. Itzykson and J.-B. Zuber, {\em Quantum Field Theory} (New York, McGraw-Hill, 1980).
%
\bibitem{12}
V. I. Nazaruk, Int. J. of Mod. Phys. E {\bf 20}, 2377 (2011); arXiv: 1101.5566 [hep-ph].
%
\bibitem{13}
V. I. Nazaruk, Eur. Phys. J. A 31, 177 (2007).
%
\bibitem{14}
V. I. Nazaruk, Phys. Lett. B {\bf 229}, 348 (1989).
%
\bibitem{15}
R. N. Mohapatra and R. E. Marshak, Phys. Rev. Lett. {\bf 44}, 1316 (1980).
%
\bibitem{16}
M. Baldo-Ceolin {\it et al.}, Z. Phys. C {\bf 63}, 409 (1994).
%
\bibitem{17}
K. S. Ganezer, presented at the International Workshop on Search for Baryon and Lepton Number Violations, Berkley, September 2007
(http://inpa.lbl.gov/blnv/blnv.htm).

\end{thebibliography}
\end{document}